\documentclass[11pt]{article}
\usepackage{moriond,epsfig}

\def\Journal#1#2#3#4{{#1} {\bf #2}, #3 (#4)}

\def\PLB{{\em Phys. Lett.}  B}
\def\PRL{\em Phys. Rev. Lett.}

\def\PRC{{\em Phys. Rev.} C}
\def\EPJC{{\em Eur. Phys. J.} C}

\def\mco{\multicolumn}

\def\be{\begin{equation}}
\def\ee{\end{equation}}
\def\bea{\begin{eqnarray}}
\def\eea{\end{eqnarray}}

\begin{document}
\vspace*{4cm}
\title{First result from the Double Chooz reactor-neutrino experiment}

\author{ T. Matsubara on behalf of the Double Chooz collaboration }

\address{Department of Physics, Tokyo Metropolitan University, Japan }

\maketitle\abstracts{
We report first results of a search for the non-zero neutrino mixing angle ${\theta}_{13}$ from the Double Chooz experiment. 
Double Chooz aims to measure the mixing angle based on $\bar{{\nu}}_{e}$ disappearance as a consequence of neutrino oscillation.  
A new generation of $\bar{{\nu}}_{e}$ detector having 10\,m$^{3}$ fiducial volume is located 1\,km from the two 4.25\,GW${}_{th}$ reactors at the Chooz Power Plant in France. 
Physics data taking has been continuing since April 2011. 
A ratio of observed-to-predicted event rate of 0.944 $\pm$ 0.016\,(stat) $\pm$ 0.040\,(syst) was obtained in 101 days of detector running. 
Analyzing both the rate and their energy spectral shape, we found sin${}^{2}$2$\theta_{13}$ = 0.086 $\pm$ 0.041\,(stat) $\pm$ 0.030\,(syst) at $\Delta m^2_{atm}$ = 2.4\,$\times$\,10$^{-3}$\,eV$^2$. 
}

\section{Introduction}

Prior to one year ago, the neutrino mixing angle ${\theta}_{13}$ had been an unknown parameter in the neutrino oscillation framework. 
The best constraint on ${\theta}_{13}$ was come from the CHOOZ reactor-neutrino experiment~\cite{bib:CHOOZ}, sin${}^{2}$2$\theta_{13}$ $<$ 0.15 (90\,\% C.L. at $\Delta m^2_{atm}$ = 2.5\,$\times$\,10$^{-3}$\,eV$^2$). 
The T2K and MINOS accelerator experiments recently reported a sign of non-zero ${\theta}_{13}$ with $\nu_{e}$ appearance~\cite{bib:T2K}$^,~$\cite{bib:MINOS}. 
First reactor experiment to present a result was Double Chooz, which reported an indication for a disappearance of reactor $\bar{{\nu}}_{e}$ in November 2011~\cite{bib:DC}. 
Other two reactor-neutrino experiments, Daya Bay and RENO, gave consistent results with higher sensitivity in early 2012~\cite{bib:DB}$^,~$\cite{bib:RENO}. 
These results supporting relatively large ${\theta}_{13}$ value, sin${}^{2}$2$\theta_{13}$ $\sim$ 0.1, potentially encourage us to measure a leptonic CP violation in near future experiments. 
This paper describes the first results of a search for the non-zero neutrino mixing angle ${\theta}_{13}$ from the Double Chooz experiment. 

Neutrino oscillation occurs as a consequence of non-zero mass and a mixing of mass eigenstates and flavor eigenstates. 
In the reactor-neutrino experiments, survival probability of $\bar{{\nu}}_{e}$ is given by: 

\begin{equation}
P_{surv} \simeq 1-\sin^22\theta_{13} \sin^2 (1.27\Delta m^2_{atm} L/E), 
\label{eq:osc}
\end{equation}

\noindent where $\Delta$$m^2_{atm}$\,(eV$^2$) is the atmospheric squared mass difference, which is precisely measured by the MINOS experiment~\cite{bib:MINOS2}. 
$L$ is the reactor-to-detector distance in meters and $E$ the $\bar{{\nu}}_{e}$ energy in MeV. 
This formula indicates the sin${}^{2}$2$\theta_{13}$ measurability by observing $\bar{{\nu}}_{e}$ disappearance at appropriate baseline. 
We therefore placed a $\bar{{\nu}}_{e}$ detector located at $\sim$1050\,m under a $\sim$300\, m.w.e. rock overburden from the two 4.25\,GW${}_{th}$ reactors at the Chooz Power Plant in France. 
In order to reduce systematic uncertainties such as neutrino flux, we will operate an identical detector located at $\sim$400\,m baseline under a $\sim$120\, m.w.e. rock overburden by 2013. 
The $\bar{{\nu}}_{e}$'s are detected through the inverse beta decay (IBD) reaction: $\bar{{\nu}}_{e}$ + p $\rightarrow$ e$^+$ + n. 
Detector based on hydrocarbon liquid scintillator provides free proton target and reacts with $\bar{{\nu}}_{e}$. 
Positron ionization and annihilation (1$\sim$8\,MeV) then creates a prompt signal. 
Neutron capture on Gadolinium ($\sim$8\,MeV) creates a delayed signal. 
The signature of IBD reaction is identified by a time coincidence of $\tau$ $\sim$ 30\,$\mu$s between those signals. 
In this reaction, the $\bar{{\nu}}_{e}$ energy can be reconstructed with prompt energy as: 
E$_{prompt}$ = E(kin.)$_{e^+}$ + 2m$_e$ $\simeq$ E$_{\bar{{\nu}}_{e}}$ - (M$_n$ - M$_p$) + m$_e$ $\simeq$ E$_{\bar{{\nu}}_{e}}$ - 0.782\,MeV. 

\begin{figure}[ht]
\begin{center}
\psfig{figure=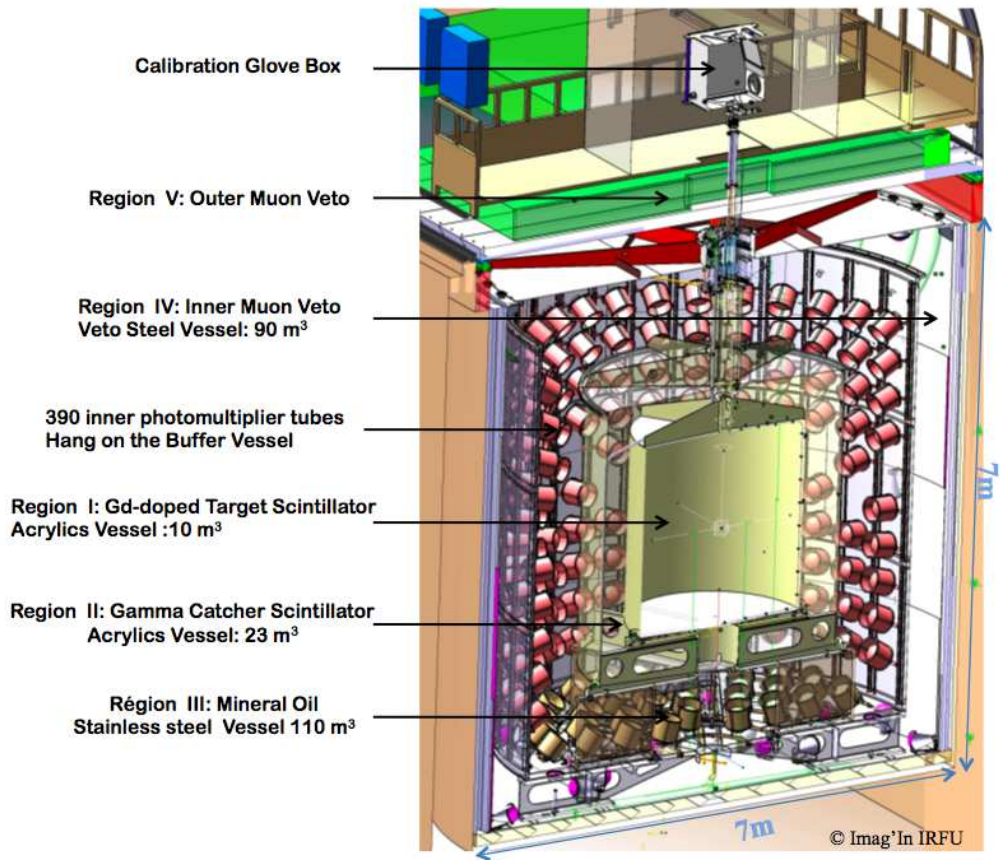,height=115mm}
\end{center}
\caption{A schematic view of the Double Chooz detector.}
\label{fig:detector}
\end{figure}

A new generation of $\bar{{\nu}}_{e}$ detector for the Double Chooz experiment (Figure~\ref{fig:detector}) consists of a main detector, an outer veto and calibration devices. 
The main detector is separated into four concentric cylindrical tanks. 
Innermost 8\,mm thick transparent acrylic vessel is called neutrino-target region. 
The region is filled with a liquid scintillator with a mixture of n-dodecane, PXE, PPO, bis-MSB and 1\,g/l Gadolinium as a beta-diketonate complex. 
This structure contains a fiducial volume for the neutrino events within 10\,m$^3$ of the target vessel. 
The composition is chosen for radiopurity and long-term stability. 
Gamma-catcher region surrounds the target vessel to detect $\gamma$-rays escaped from neutrino-target region. 
It consists of 12\,mm thick transparent acrylic vessel containing 22.3\,m$^3$ of liquid scintillator. 
Buffer region is located outside of the gamma-catcher region. 
The region is filled with 110\,m$^3$ of mineral oil to shield $\gamma$-rays from the Photo Multiplier Tubes (PMT) and rocks. 
Total 390 10-inch low-radioactive PMTs are equipped at the inner wall of stainless steel buffer tank to observe light from the inner volumes. 
These three layers constitute the inner detector (ID). 
The ID is surrounded by the optically separated region of the inner veto (IV) equipped with 78 8-inch PMTs. 
90 m$^3$ liquid scintillator is filled in a steel tank for muon veto and shielding of spallation neutrons produced outside of the detector. 
The detector is surrounded by 150\,mm of demagnetized steel to reduce a contamination due to external $\gamma$-rays. 
Outer veto system lies on main detector, which is not used for this analysis. 
The main detector is read out with 500\,MHz flash-ADC electronics with customized firmware and a deadtime-free acquisition system.

\section{Neutrino prediction and data analysis}

The analysis is based on 101.5 days of data including 16 days with one reactor off and one day with both reactors off. 
The expected number of $\bar{{\nu}}_{e}$'s are predicted as: 

\begin{equation}
N^{exp}_{\nu}(E, t) = \frac{N_p\epsilon}{4\pi L^2} \times \frac{P_{th}(t)}{\langle E_f \rangle} \times \langle \sigma_f \rangle, 
\label{eq:osc2}
\end{equation}

\noindent where $N_p$ is the number of protons in the detector, $\epsilon$ the detection efficiency, $P_{th}(t)$ the thermal power of reactor, $\langle E_f \rangle$ the mean energy per fission and $\langle \sigma_f \rangle$ the mean cross-section per fission. 
Mean energy per fission and associated errors are evaluated with two reactor simulation codes, MURE~\cite{bib:MURE} and DRAGON~\cite{bib:DRAGON}. 
Mean cross-section per fission and associated errors are estimated with updated reference spectra~\cite{bib:Spectra}. 
Neutrino prediction is based on Bugey4 measurement~\cite{bib:Bugey4} as an anchor point of the mean cross-section with correction to Chooz reactor. 
Total systematic uncertainty on reactor is estimated to be 1.8\,\%. 

Energy measurements for data analysis are based on the total charge collected by the PMTs. 
The energy is reconstructed scaling the total charge by a constant corresponding to $\sim$200\,p.e./MeV, which is adjusted with the 2.2 MeV energy peak of neutron capture on H at the target center. 
Our Monte Carlo simulation based on GEANT4 is used to calculate the detector response.  
Comparison between actual and simulated calibration data has been done to correct the simulation and estimate an associated uncertainty to be 1.7\,\%, using two parametric functions with respect to energy and position. 

\begin{figure}[ht]
\begin{center}
\psfig{figure=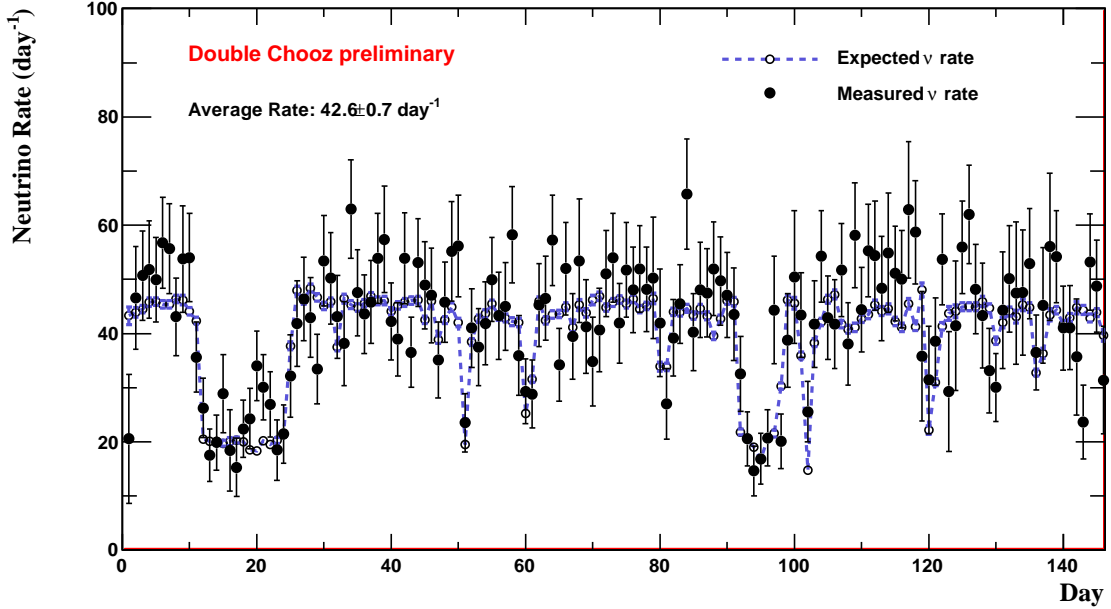,height=90mm}
\end{center}
\caption{Observed and expected neutrino event rate.}
\label{fig:neutrino_rate}
\end{figure}

We applied the following criteria to select $\bar{{\nu}}_{e}$ candidates. 
Triggers within a 1000\,$\mu$s after a cosmic muon crossing the ID or IV are vetoed to suppress spallation neutrons and cosmogenic backgrounds. 
This requirement is followed by five selections: (1) a cut rejecting events caused by sporadically glowing PMT bases, resulting in light localized to a few PMTs and spread out in time: Q$_{max}$/Q$_{tot}$ $<$ 0.09 (0.06) for the prompt (delayed) energy and rms(t$_{start}$) $<$ 40\,ns, where Q$_{max}$ is the maximum charge recorded by a single PMT and rms(t$_{start}$) is the standard deviation of the times of the first pulse on each PMT; (2) 0.7\,MeV $<$ E$_{prompt}$ $<$ 12.2\,MeV; (3) 6.0\,MeV $<$ E$_{delayed}$ $<$ 12.0\,MeV; (4) 2\,$\mu$s $<$ $\Delta t_{e^+n}$ $<$ 100 $\mu$s; (5) a multiplicity cut to reject correlated backgrounds defined as no additional valid trigger from 100\,$\mu$s preceding the prompt candidate to 400\,$\mu$s after it. 
In total, 4121 neutrino candidates were observed, which is equal to 42.6 $\pm$ 0.7 events/day on average (Figure~\ref{fig:neutrino_rate}). 

Backgrounds mimicking the time coincidence have been estimated as follows. 
Accidental background caused by uncorrelated coincidence, for example, prompt event from radioactivity and delayed event from cosmogenic neutron capture. 
This background rate is estimated by sequentially shifted off-time window, leading 0.33 $\pm$ 0.03 events/day. 
Fast neutron induced by muon traveling the rock can interact in the detector producing recoil proton. 
Thermalized and captured neutron accompanying the recoil proton then mimics the IBD events. 
We estimated the rate to be 0.83 $\pm$ 0.38 event/day by modifying the E$_{prompt}$ selection criteria to be 12.2\,MeV $<$ E$_{prompt}$ $<$ 30\,MeV and extrapolating to the signal region assuming flat energy spectrum. 
Spallation product of $^9$Li induced by energetic muons emits n and $\beta$, simulating IBD event. 
This background is studied by time distribution between a muon energy deposition $>$ 600\,MeV and neutrino candidates. 
Fitting the time distribution by a flat component and an exponential with the $^9$Li lifetime results in 2.3 $\pm$ 1.2 events/day. 
We account for a shape uncertainty between some decay branches. 

We had a chance to take data with both reactors off for $\sim$24\,hours. 
It was an unique opportunity to validate our background estimation. 
Two candidates in neutrino energy window following high energy muon were observed, which are compatible with candidates of $^9$Li background. 
This result is consistent with the estimated number of BG events (Total 3.46 $\pm$ 1.26 events/day). 

\section{Oscillation analysis}

\begin{table}[ht]
\caption{Summary of uncertainties}
\vspace{0.4cm}
\begin{center}
\begin{tabular}{|c|l|c|c|}
\hline
\mco{2}{|c|}{Source} & \mco{2}{|c|}{Uncertainty w.r.t signal}\\ \hline
\mco{2}{|l|}{Statistics} & \mco{2}{|c|}{1.6\,\%}\\ \hline
 & Bugey4 measurement & 1.4\,\% & \\
 & Fuel composition & 0.9\,\% & \\
 & Thermal power & 0.5\,\% & \\
 Reactor& Reference spectra & 0.5\,\% & 1.8\,\%\\ 
 & Energy per fission & 0.2\,\% & \\ 
 & IBD cross section & 0.2\,\% & \\ 
 & Baseline & 0.2\,\% & \\ \hline
 & Energy response & 1.7\,\% & \\ 
 & E$_{delay}$ containment & 0.6\,\% & \\ 
 & Gd fraction & 0.6\,\% & \\
Detector& $\Delta t_{e^+n}$ & 0.5\,\% & 2.1\,\%\\
 & Spill in/out & 0.4\,\% & \\ 
 & Trigger efficiency & 0.4\,\% & \\ 
 & Target H & 0.3\,\% & \\ \hline
 & Accidental & $<$ 0.1\,\% & \\ 
Background& Fast neutron & 0.9\,\% & 3.0\,\%\\ 
 & $^9$Li & 2.8\,\% & \\ \hline
\end{tabular}
\end{center}
\label{tab:uncertainties}
\end{table}

\begin{figure}[ht]
\begin{center}
\psfig{figure=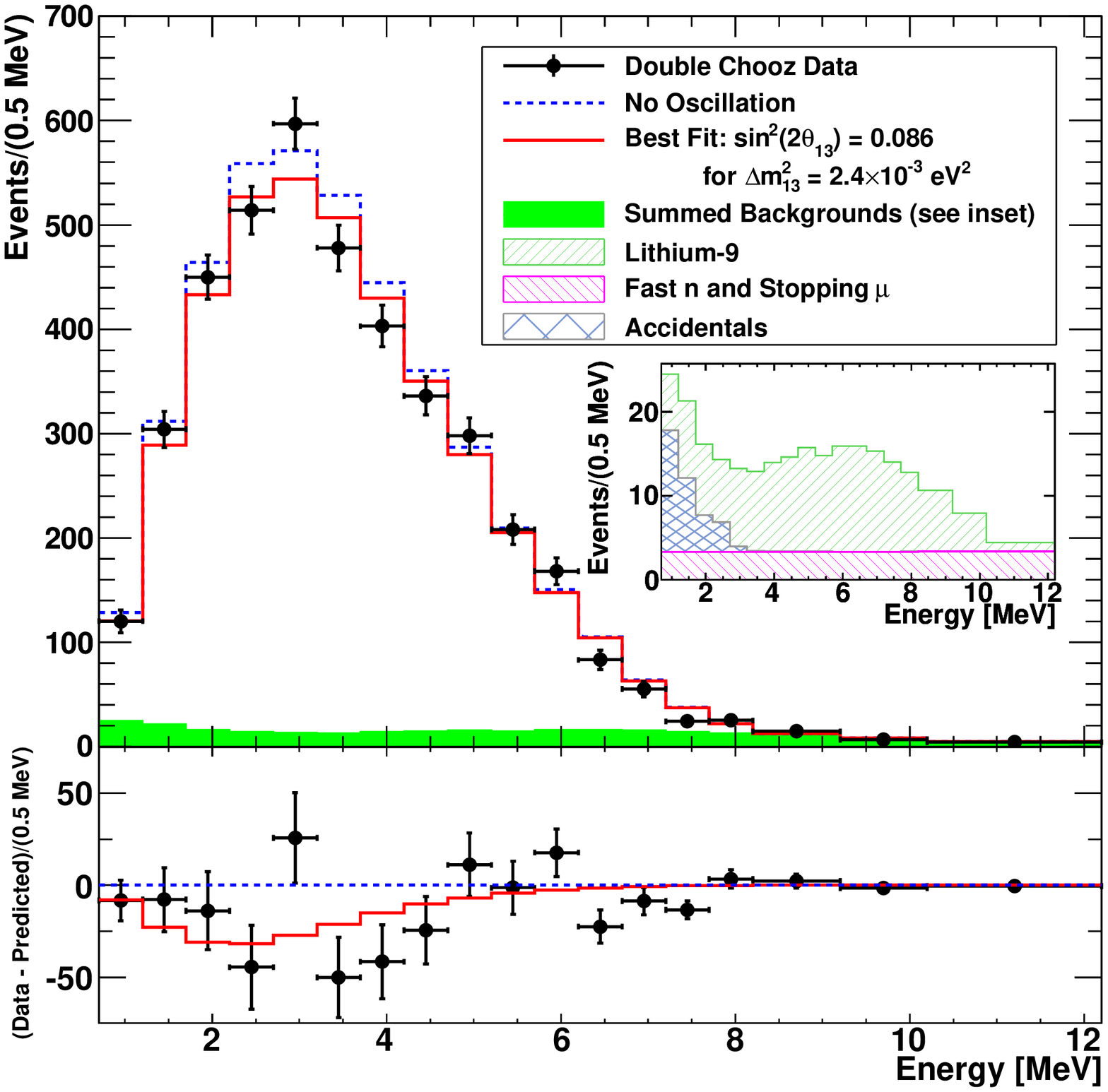,height=85mm}
\end{center}
\caption{Top: Observed and predicted positron energy spectra for the best-fit sin${}^{2}$2$\theta_{13}$ and no-oscillation expectation with stacked background spectrum. Bottom: Difference between data and no-oscillation (dots) and difference between the best-fit sin${}^{2}$2$\theta_{13}$ and no-oscillation expectation (line). }
\label{fig:fit result}
\end{figure}

\begin{figure}[ht]
\begin{center}
\psfig{figure=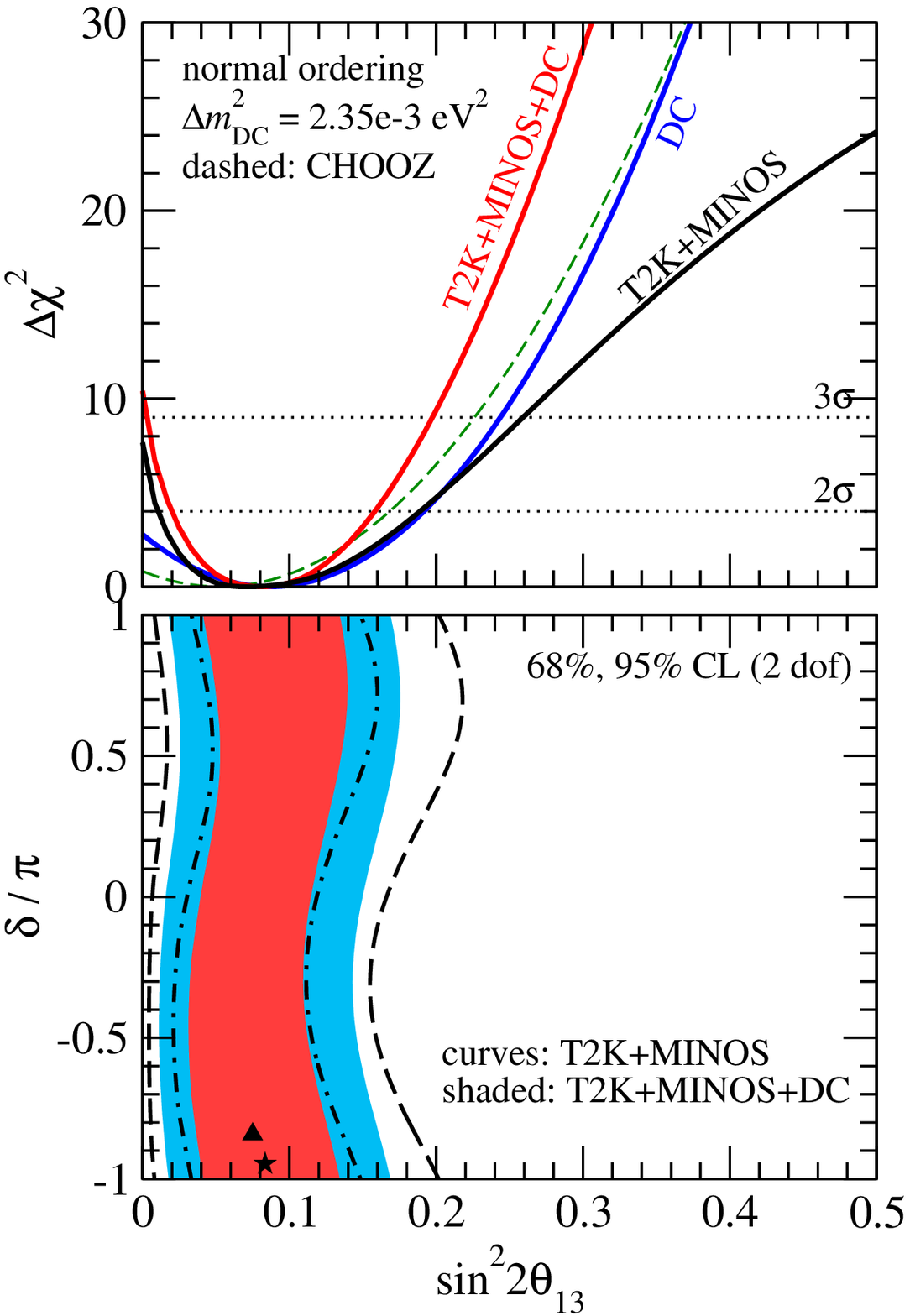, height=115mm}
\caption{Combined analysis with Double Chooz, T2K and MINOS assuming normal mass ordering. The $\Delta \chi^2$ distribution for sin${}^{2}$2$\theta_{13}$ (Upper plot) and the allowed region in $\delta$ vs  sin${}^{2}$2$\theta_{13}$ plane (Lower plot), respectively.}
\label{fig:combined_result}
\end{center}
\end{figure}

Oscillation analyses have been done based on $\chi^2$ estimator. 
Uncertainties used for the oscillation analyses are summarized in Table~\ref{tab:uncertainties}.  
A ratio of observed-to-predicted events of 0.944 $\pm$ 0.016\,(stat) $\pm$ 0.040\,(syst) was observed, 
corresponding to sin${}^{2}$2$\theta_{13}$ = 0.104 $\pm$ 0.030\,(stat) $\pm$ 0.076\,(syst) at $\Delta m^2_{13}$ = 2.4\,$\times$\,10$^{-3}$\,eV$^2$. 
The analysis can be improved by using spectral information. 
The positron spectrum is divided into 18 variably sized energy bins from 0.7\,MeV to 12.2\,MeV. 
In order to introduce bin-to-bin correlations, we use four covariance matrices to include uncertainties on statistics, reactor, detector and background spectral shape. 
Shape difference between signal and background appears in the region from 8 to 12.2\,MeV, reducing the uncertainties due to correlated backgrounds. 
Analyzing both the rate and their energy spectral shape, we found sin${}^{2}$2$\theta_{13}$ = 0.086 $\pm$ 0.041\,(stat) $\pm$ 0.030\,(syst) with $\chi^2$/n.d.f. = 23.7/17. 
Observed and predicted positron energy spectra for no oscillation and the best-fit sin${}^{2}$2$\theta_{13}$ including background are shown in Figure~\ref{fig:fit result}. 
No oscillation hypothesis is ruled out at the 94.6\,\%\,C.L., which can be interpreted as an indication of the non-zero ${\theta}_{13}$. 
This result is compatible with follow-up results from the Daya Bay and RENO experiments. 
A combined analysis with the T2K and MINOS accelerator experiments on $\theta_{13}$ and CP violation phase $\delta$ vs $\theta_{13}$ plane for normal mass ordering is presented as shown in Figure~\ref{fig:combined_result}. 

\section{Conclusion}

Double Chooz started a search for the non-zero neutrino mixing angle ${\theta}_{13}$ using the new generation of $\bar{{\nu}}_{e}$ detector since April 2011. 
A ratio of observed-to-predicted events of 0.944 $\pm$ 0.016\,(stat) $\pm$ 0.040\,(syst) was observed in 101 days of detector running. 
We found sin${}^{2}$2$\theta_{13}$ = 0.086 $\pm$ 0.041\,(stat) $\pm$ 0.030\,(syst) at $\Delta m^2_{atm}$ = 2.4\,$\times$\,10$^{-3}$\,eV$^2$, based on rate and spectral shape information. 
The no oscillation hypothesis is excluded with 94.6\,\% C.L., which can be interpreted as an indication of the non-zero ${\theta}_{13}$.

\end{document}